# A Description Driven Approach for Flexible Metadata Tracking

Andrew Branson (1), Jetendr Shamdasani (2), Richard McClatchey (3)

*University of the West of England*
*Coldharbour Lane, BS16 1QY, Bristol, England*

(1)*EMail: Andrew.Branson@cern.ch*

(2)*Email: Jetendr.Shamdasani@cern.ch*

(3)*EMail: Richard.McClatchey@cern.ch*

## ABSTRACT

Evolving user requirements presents a considerable software engineering challenge, all the more so in an environment where data will be stored for a very long time, and must remain usable as the system specification evolves around it. Capturing the description of the system addresses this issue since a "description-driven" approach enables new versions of data structures and processes to be created alongside the old, thereby providing a history of changes to the underlying data models and enabling the capture of provenance data. This description-driven approach is advocated in this paper in which a system called 'CRISTAL' is presented. CRISTAL is based on description-driven principles; it can use previous versions of stored descriptions to define various versions of data which can be stored in various forms. To demonstrate the efficacy of this approach the history of the project at CERN is presented where CRISTAL was used to track data and process definitions and their associated provenance data in the construction of the CMS ECAL detector, how it was applied to handle analysis tracking and data index provenance in the neuGRID and N4U projects, and how it will be matured further in the CRISTAL-ISE project. We believe that the CRISTAL approach could be invaluable in handling the evolution, indexing and tracking of large datasets, and are keen to apply it further in this direction.

Keywords: CMS, Metadata, Schema Evolution, Computer Science, Provenance, Semantics

## INTRODUCTION

Many organizations operate in environments which dictate changes at an unforeseeable rate and force evolution of system requirements. If their periods of system design are such that their supporting technologies may evolve or are not mature at the outset, then the design process becomes considerably more difficult. In these circumstances the design must also need to be able to evolve in order to react to changing technologies and to ensure traceability between the fluid design and the evolving system specification. This is how we define the use of what is known in the area of Computer Science as *Provenance* [1]. This is one aspect of the data preservation problem, where developers need to control how their application progresses over time and to be able to revert back to older versions of their models if required. This is how we define the *evolution of models*. The strength to this approach is that developers can query past models and see how models have changed/evolved during the course of system design. Everything is stored and archived so that these models can be *queried* and *reused* at a later date if required.

In this paper we outline how such an approach has been facilitated in practice by implementing what we term a 'description-driven' system [2]. Essentially the description-driven approach involves identifying and abstracting, at the outset of design, all the crucial elements (such as business objects, processes, lifecycles, goals, agents and output) in the system under consideration and creating high-level descriptions of these elements which are stored in a model, dynamically modified and managed separately from their instances.

The next sections describe the motivation for this research in terms of a practical example (called CRISTAL) based around large-scale engineering, and how the system evolved into a flexible fully traceable data indexing catalogue, which can store provenance on data files and analyses.

## THE NEED FOR DESCRIPTIONS

The principal motivation in carrying out this research was to address the requirements of a community of scientific researchers working at a number of locations concurrently over extended timescales on the design of a large experiment. The example studied was that of the Large Hadron Collider (LHC) accelerator project at CERN [3] which reached its final phase of construction and testing in 2009. The Compact Muon Solenoid [4] is one of the four main experiments of the LHC; it contains the so-called Electromagnetic Calorimeter (ECAL) detector comprising an enveloping array of about 70,000 individual single crystals of lead tungstate (PbWO4) that can measure the exact location and energy of high energy electromagnetic particles from collisions in the heart of the CMS. Its design and construction were carried out between 1995 and 2008, and it was a key detector in finding the Higgs Boson.

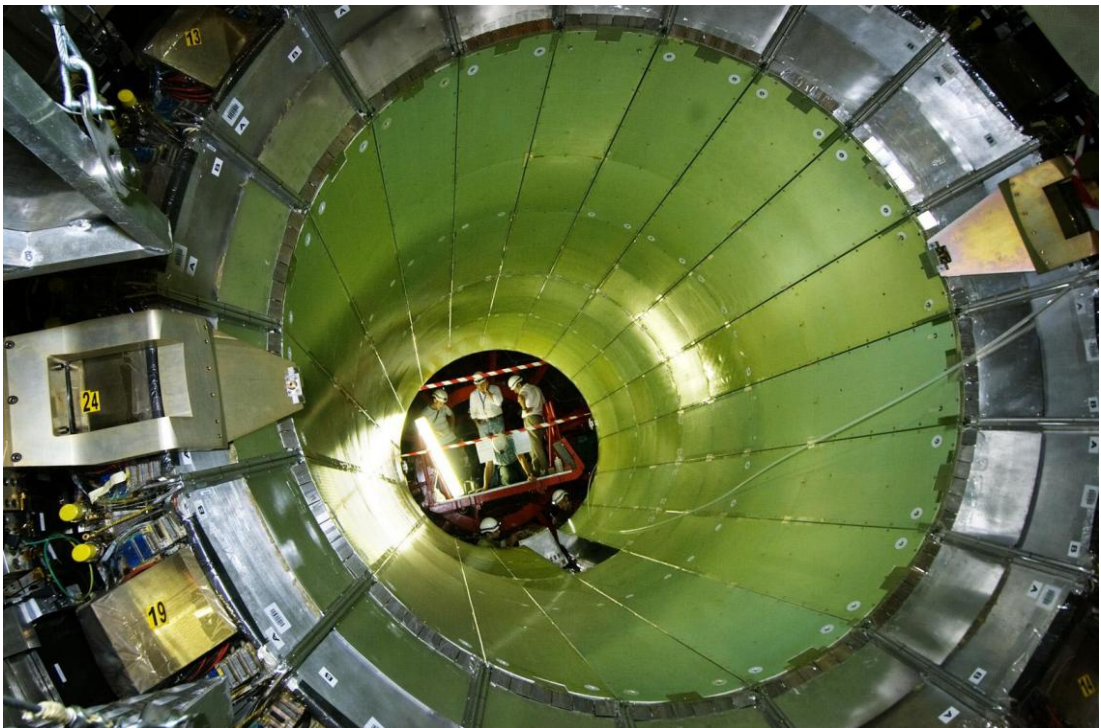

Figure 1: Insertion of the last CMS ECAL supermodules in the ECAL Barrel. © CERN July 2007

Each element of the CMS ECAL detector has been constructed from a large number of precision parts produced and assembled during the last decade by centers distributed worldwide. Each constituent part needed its physical characteristics to be measured, producing considerable quantities of data, prior to its assembly and integration in the experimental area at CERN. Furthermore each part must be located in the final assembly. The data collected during this phase is needed to construct the detector and its later calibration and operation. Any system for managing this data must provide traceability of composite parts and processes and be evolvable over time. The initial logic for ECAL construction changed considerably throughout the project, while the data needed to be gathered, managed and curated for future use. The specification of the ECAL took more than a decade during which time many fundamental decisions impacting its design, detector components' physical characteristics and data formats were made, so that any database system used to track those changes had to be able to handle frequent changes and thus had to be very flexible in design.

A research project, entitled CRISTAL (Cooperating Repositories and an Information System for Tracking Assembly Lifecycles) [5] was established, using pure object-oriented computing technologies where possible, to facilitate the management of the engineering data collected at each stage of construction of the detector. CRISTAL is a distributed product data and workflow management system which uses a database for its repository, a multi-layered architecture for its component abstraction and dynamic object modelling for the design of the objects and components of the system. These techniques are critical in handling the complexity of such a data-intensive system and to provide the flexibility to adapt to the changing production scenarios typical of any research production system.

Designing a system to provide product data management, workflow tracking and change management to an agreed set of user requirements is a challenging task and there have been many previous projects aimed at this. However in a research environment, where new materials and associated electronics are being evaluated, tested and developed in a research-led manner it is simply not possible to specify at the outset of a 10-12 year design and development cycle precisely the form of the final detector or the production process for its construction. However researchers still need to track all the steps of this process from early prototyping to completion, which entailed the capturing of design versions with the storage and management of all data associated with instantiations of those versions.

The CRISTAL system that has been developed to address these requirements has been based on a so-called description-driven approach in which all logic and data structures are "described" by meta-data, which can be modified and versioned online as the design of the detector changes. A description-driven system (DDS) architecture, as advocated in this paper and previously in (Estrella et al, 2003 [5]) is an example of a reflective meta-layer architecture. DDSs make use of meta-objects to store diverse domain-specific system descriptions (such as items, processes, lifecycles, goals, agents and outcomes) which control and manage the lifecycles of instances or domain objects. As objects, reified system descriptions of DDSs can be organized into libraries conforming with frameworks for modelling of languages in general, and to their adaptation for specific domains.

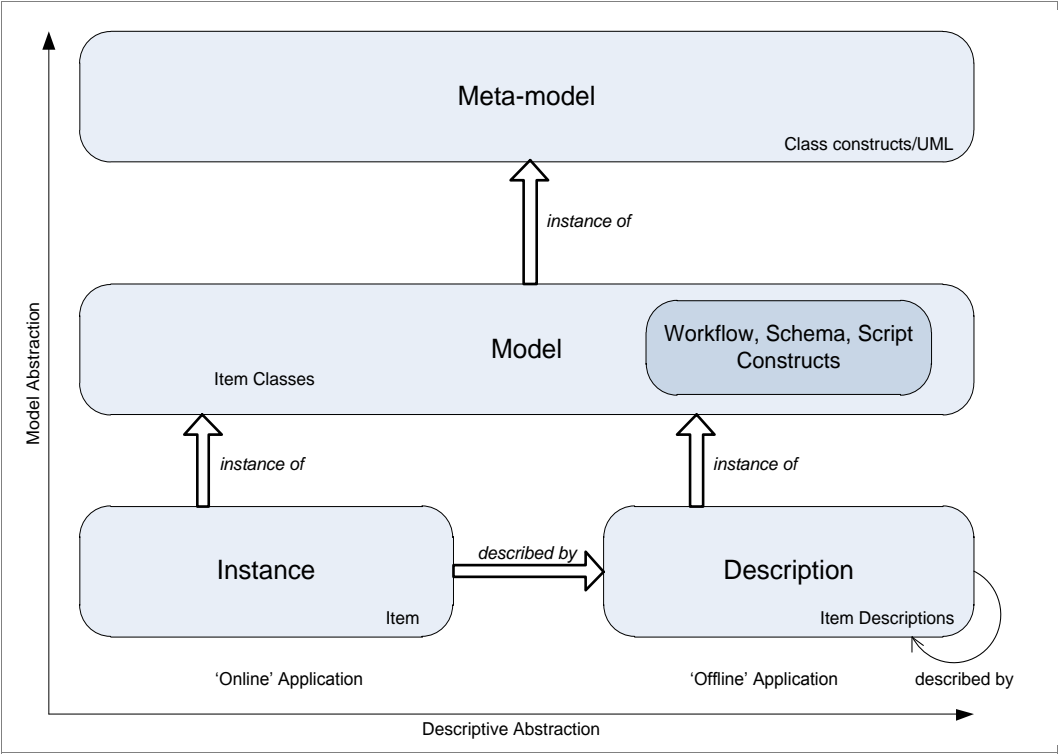

Figure 2 : Model vs Descriptive Abstraction in CRISTAL

The meta-data and the instantiated elements of data are stored in the database; the evolution of the design is tracked by versioning the changes in the meta-data over time. Thus description-driven systems make use of meta-objects to store domain-specific system descriptions that control and manage the lifecycles of domain objects (that is instances of the meta-object definitions). The separation of the descriptions from their instances allows them to be specified and managed also to evolve independently and asynchronously. This separation is essential in handling the complexity issues facing many web-based computing applications and facilitates interoperability, reusability with system evolution. Separating descriptions from their instantiation allows new versions of defined objects (and in turn their descriptions) to coexist with older versions. The reader is directed to previous publications on DDS for further background [5-6].

At the outset of CRISTAL an approach was followed that enabled both ECAL parts (with the metadata of their specifications) and the activities (with the metadata of their specifications) that were carried out on the parts to be saved side-by-side in a structured database. In this way as the scientists developed their research ideas we were able to capture each design version and those parts and activities that were processed for that version. The production version of CRISTAL was designed to satisfy a set of requirements covering both the CMS ECAL needs and the needs of commercial BPM users. This led to a generalization of the concept of 'parts' to that of 'items' which were intended to be applicable to describe any 'item' in the real world as an example an element of any business process. The intention was to enable developers to define the objects central to the operation of any business and to drive the design process from the standpoint of how those objects ('items') change over time. The production version of the CRISTAL kernel was spun off as a BPM application for industry to the French company Agilium in 2003 [7]. Agilium is still going strong, and another company, Technoledge [8], has since been created to focus more on the data management side of CRISTAL.

## THE CRISTAL MODEL

This section describes (briefly) the internal workings of CRISTAL and what its model contains. It should be noted that this is just a brief description. For a full description of the inner workings of CRISTAL please see [9]. CRISTAL is, in essence, an application server that abstracts all of its business objects into workflow-driven, version-controlled 'Items' which are instantiated from descriptions stored in other Items (see Figure 2) and are managed on-the-fly for target user communities.

Items contain:

- Workflows, that comprise of Activities to be executed by Agents (either human users or mechanical/ computational agents via an API), which then generate:
- Events that detail each change of state of an Activity. Completion events generate data, stored as:
- Outcomes which are XML documents from each execution, for which:
- Viewpoints refer to particular versions (e.g. the latest version or, in the case of descriptions, a particular version number).
- Properties are name/value pairs that name and type items, they also denormalize collected data for more efficient querying, and
- Collections that enable items to be linked to each other.

Item Description Items hold the templates for new Items, and also dictate their type (see Figure 3). They are themselves also Items (and thus the two can be treated in the same manner), holding the description data as XML outcomes managed through workflow activities. Workflow and Property descriptions are stored as XML serialized objects. Collection Descriptions are themselves Collections, pointing to other Item Descriptions. Outcome Descriptions contain XML Schema documents which are used to validate submitted outcomes and aid in data collection, for instance to generate data entry forms in the stock GUI for the end users. Also included in the descriptions are Scripts - snippets of code invoked by workflows either during a change of Activity state to enact consequences of the execution such as updating a Property or changing a Collection, or to assess conditional splits in the Workflow based on existing data. Scripts are given the full context of the Item and Activity in which they are running, so they may invoke other Activities (which may in turn trigger other scripts), or even navigate Collections to interact with other Items. They are stored in an XML Outcome along with declarations of their input parameters and output.

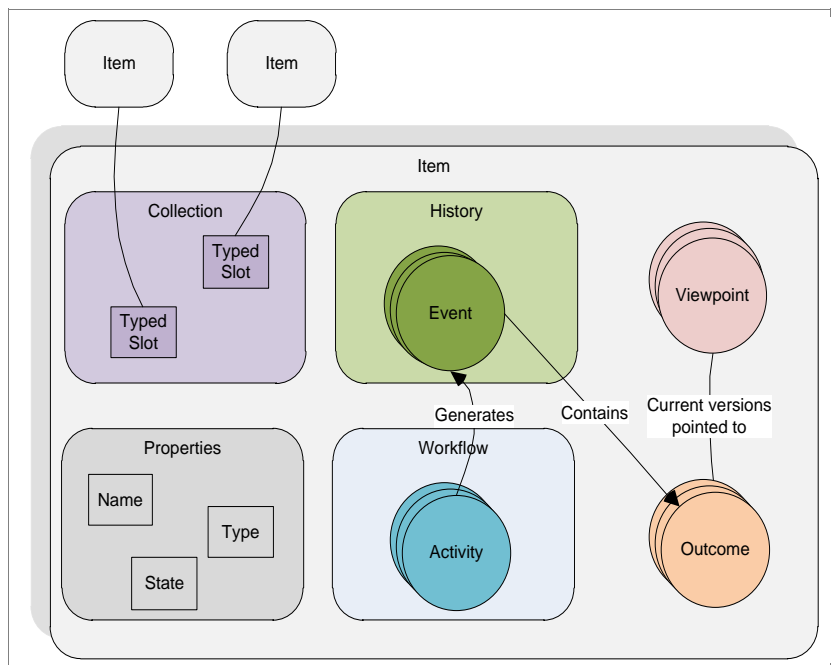

Figure 3: The components of an Item in CRISTAL

Writing to the CRISTAL object model is impossible from a client process except through activity execution, thus providing full traceability of the system. Ordinary activities only create new Events and Outcomes, and modify Viewpoints, so when a script needs to modify some other part of the model it must invoke special 'Predefined Steps' which are Activities that contains additional logic for modifying the Item's Properties, Collections or directory entries. These Predefined Steps are hard-coded in the kernel and do not often change, making their presence in an Item's history reliably interpretable.

The aim of this rigidity of write control is to require the design of the lifecycle of each Item type to explicitly define the full behaviour of that Item. We see this as a return to the principles of the object modelling that many modern languages and platforms have fallen behind on in the name of rapid prototyping and cutting down code volume, whereas a properly designed metamodel should achieve those without sacrificing object orientation.

The CMS ECAL was constructed from thousands of similar parts, all needing characterizing and assembling in an optimal configuration based on sets of detailed measurements. Every component part was registered as an Item in the CRISTAL database, each with its barcode as an identifier. Each part had a type, which functioned as the item description, and was linked to the workflow definition that each instance would follow in order to collect its data and mount sub-parts. All collected data and assembly information were stored as outcomes attached to events, and therefore, the entire history of every interaction with the application was recorded. The result was a set of items representing the top level components of the detector which contained levels of substructure, all with their full production history with all collected and calculated production data attached in the correct context.

## THE DESCRIPTION-DRIVEN DEVELOPMENT LIFECYCLE

Conventional software development separates the specification phase from the construction and implementation phases. However, when the design is evolving as a result of changing user needs, the development process must be reactive and necessarily iterative in nature. The new requirements from the users need to be implemented by the developer in an incremental fashion so that the new results can be assessed and further changes to the design requested, if needed. CRISTAL allows the user to directly verify the business object workflow design, so the normal progression through implementation and testing can be short-circuited. In other words the users can visualize the overall process to be captured in terms of their own recognizable world objects; this greatly simplifies the analysis and (re-)design process. It is relatively easy for professional users to understand the workflow system in CRISTAL, and the nature of XML based data; these both can be detailed by the application maintainer sufficiently accurately in collaboration with the user or may even be drawn by an expert user directly.

The application logic that needs to be executed during the workflow will have its functionality conveniently broken down along with the activities. It is then simple to import these definitions into the system where it can be immediately tested for feedback to the users. Improvements can thereby be quickly performed online, often by modifying the workflow of one test item, which then serves as a template for the type definitions. Items subject to the improvements can co-exist with items generated earlier and prior to the improvement being made and both are accessed in a consistent, reusable and seamless manner. All this can be done without recompiling a single line of code or restarting the application server, providing significant savings in time and enables the users to work in an iterative and reactive manner that suits their research, often adopting rapid prototyping principles.

In our experience, the process of factoring the lifecycle and dataset of the new item type into activities and outcomes helps to formalize the desired functionality in the user's mind; it becomes more concrete - avoiding much of the vague and often inconclusive discussion that can accompany user requirements capture. Because it evolved from a production workflow specification driven by user requirements, rather than a desire simply to create a 'workflow programming language', CRISTAL's style of workflow correlates more closely to the users' concept of the activities required in the domain item's lifecycle. The degree of granularity can be chosen to ensure that the user feels it provides sufficient control, with the remaining potential subtasks rolled up into a single script. This is one important aspect of the novel approach adopted during CRISTAL development that has proven of benefit to its end-user community. In practice this has been verified over a period of more than 10 years use of CRISTAL at CERN and by its exploitation as the Agilium product across many different application domains (see discussion in the later sections).

When working with users certain questions need to be asked so that requirements can be gathered and a methodology had to be developed so that users can work with the CRISTAL system in a standard manner. Some that we identified at the end of the process are the following:

1. Firstly, the design of the normal workflow of a new item type is outlined on paper with the user. As each activity is identified, any data collected during its execution is also written down. Any computation that needs to be performed is broken down into individual scripts.

2. The standard CRISTAL user interface is normally usable for executing virtually any activity, but it is rather abstract. So then the question is asked, will this suffice for this workflow, or is a simpler custom interface required? Are the end users familiar with CRISTAL or not?

3. Next, any exceptional paths followed in the user workflows or patterns are identified. Common ones are implemented with conditional splits in the workflow. If there is too much branching, there is a risk that the user will be overwhelmed or will spend too much time dealing with the choices. Others can be omitted until they get more precisely defined through experience.

4. Once the workflow design is complete, the new types can be created in CRISTAL and one test item can be instantiated. Essential changes are implemented in the instance workflow, and the data is cleared and the workflow restarted as required. Once this is stable, the changes can be copied back to the description, and normal production may (re-)start.

For the invocation of custom user application logic, CRISTAL includes a generic user-code' process, which can log in and execute jobs generated by activities as they change state. This will execute any scripts associated with transitions as part of the execution process, but the process can also be extended to execute other code as necessary. Thus, there are three methods available for deployment of application logic:

1. As a CRISTAL script: This approach is the most desirable, since it supports versioning alongside the rest of the descriptions. Although CRISTAL supports many scripting languages through the Bean Scripting Framework, including Javascript, Python and Perl, there is still significant migration required to convert existing compiled logic.

2. Using a custom user-code agent as a wrapper: This was the method used to migrate existing CRISTAL V1.0 logic, written in Java, to CRISTAL V2.0. This has the advantage of requiring little or no modification of the existing code once the wrapper is complete, but modifications require redeployment of server code.

3. A mixture of the two: scripts to invoke external code.

Once the prototyping has been completed the domain application can finally go into production. From this point onwards data entered into the system must be preserved and any changes to the domain application, defined entirely in descriptions, must maintain backwards compatibility with it. The designer's work is completed, and the maintainer's begins. When an item design goes into production, the chances are that it will have to be modified due to either changing user requirements or inaccurate/unspecified initial requirements. These new requirements may manifest themselves to the domain application maintainer in a number of different ways.

Sometimes, and most conveniently, the users will present a modified workflow design directly, which after examination for possible inconsistencies can be directly implemented in the system. A user who is well versed in workflow design may even ultimately be given the rights to modify the workflow themselves. However, more often the user will request some modification by the application maintainer to the item's data or workflow, to enable such a state that is not possible with the current design. In this case, the application maintainer must assess the likelihood of this requested intervention happening again. If they think it will be a rare occurrence, then it can be quickly fixed by exceptionally manipulating the item's data directly, although care must be taken to make sure that the altered Item is left in a consistent state, limiting the number of people who can perform the fix to domain developers who have a deep knowledge of the Item design. Scripts can be written to apply an identical change to many items at once. If the original design was clearly wrong, then the workflow description can be modified for future items and existing items can be migrated if possible.

To provide traceability, all changes of state in CRISTAL activities are themselves logged as events in the histories of items. Although it is possible to write a script in such a way as to write to the data storage directly, this method will leave no trail for later debugging and it must also be run in the server process itself (the only place with write access to the data stores). Consequently scripts interact with their items through activity execution (most usefully through the system predefined activities).

It is worth noting that this is an iterative arrangement – the application maintainer in one situation is himself a user of description development tools maintained by the domain maintainer above, who deploys and maintains those tools on top of the kernel itself. To draw an analogy with the world of object orientation, the domain expert writes the classes with which (s)he then defines different types. In ECAL, the domain consists of Products and Orders, and the different detector components and the transfer requests to move them around are defined from those Orders.

The Event model in CRISTAL is very powerful and it provides versioning of all data and transformations to models conducted within CRISTAL. Everything is stored and logged for later usage via an XML DB and can be queried and accessed at a later date via a custom schema. Therefore, nothing is lost in CRISTAL and the evolution of models can be shown over time. The user is first given access to a particular viewpoint of an item (the last version) and this version can be changed to an older version as well. Also, the latest version is not lost and it kept for later if required to be used. Therefore, from a provenance standpoint the full trace of what has happened to an item from inception to production execution is available so that application developers and maintainers can see what has occurred to an Item and ascertain why certain changes have been applied.

## AFTER THE CMS ECAL CONSTRUCTION

At the beginning of the project, CRISTAL was actively developed by The University of the West of England (UWE, United Kingdom), the Centre for European Nuclear Research (CERN, Switzerland) and the Centre National de la Recherche Scientifique (CNRS, France). This consortium is currently moving towards an open source release of the CRISTAL kernel and this will occur early 2014. This section describes work that is currently ongoing since the end of the construction of the CMS ECAL. There is currently sizeable interest in CRISTAL and what it can provide for both the commercial and research / academic worlds.

The CRISTAL system is commercially exploited by the Agilium company (now M1i, Annecy, France) for the purpose of supporting business process management (BPM) and the integration and cooperation of multiple business processes especially in business-to-business applications. The product, itself called Agilium, addresses the harmonisation of business processes by the use of a CRISTAL database so that multiple potentially heterogeneous processes can be integrated with each other and have their workflows tracked in the database. Agilium also integrates the management of data coming from different sources and unites BPM [10] with business activity management (BAM) [11] and Enterprise Application Integration (EAI) [12] through the capture and management of their designs in the CRISTAL database. Using the facilities for description and dynamic modification in CRISTAL, Agilium is able to provide modifiable and reconfigurable business process workflows, not attached to a particular physical object, but existing as a business process alone. It uses the description-driven nature of the CRISTAL-Agilium models to act dynamically on process instances already running and can thus intervene in the actual process instances during execution. These processes can be dynamically (re-)configured based on the context of execution without compiling, stopping or starting the process and the user can make modifications directly and graphically of any process parameter. Thus the Agilium system aims to provide the level of flexibility for organisations to be agile in responding to the ongoing changes required by cyber-enterprises. Further details of the Agilium product can be found at [7].

The University of the West of England (UWE) has focused on enhancing the modelling capabilities of the core CRISTAL software and applying them to other domains. A limiting factor in early deployments of CRISTAL was the backend storage of the CRISTAL objects - the choice was to either serialize them to XML on a filesystem or map their data onto relational database tables. Querying either was unwieldly at best, and impossible at worst. In recent years, XML databases have matured into stable, schemaless databases with performant query results and manageable deployment requirements. In using an XMLDB behind CRISTAL, we have found a powerful combination of flexible schemaless data storage, coupled with schema based data validation on write, making XQueries developed against runtime defined schemas efficient and easier to compose.

Recently CRISTAL has been studied as a 'provenance data management system' or a system that is used to manage the history of data and workflows and their usage over time in a scientific environment. In the neuGRID project and its follow-up N4U, supported by the EC 7th Framework Programme since 2009, we have been implementing Provenance and Querying Services using a CRISTAL kernel for the purposes of supporting neuroscientists in their studies of Alzheimer's disease across Europe (see [13] and [14]). The use of CRISTAL as the Provenance Service database and engine has enabled neuroscientists to track their complex image datasets and analyses over time and to collaborate together in teams with fully versioned workflows and datasets. The flexible schema model we have is more than capable of integrating metadata from many different imaging sets, of whatever format, together in one database (coined the 'Data Atlas') which allows neuroscientists to search for specific image properties for their analyses across all available datasets.

## CONCLUSIONS AND FUTURE PLANS

Describing a proposed system explicitly and openly from the outset of the project enables the developer to change aspects of it responsively as users' requirements evolve. This enables seamless transition from version to version with (virtually) uninterrupted system availability and facilitates full traceability throughout the system lifecycle. Following the fundamental principles of object-oriented design our model-driven approach encourages the reuse of code, configuration data and scripts/methods. Indeed, the description-driven design approach takes this one step further and provides reuse of meta-data, design patterns [15] and maintenance of items and activities (and their descriptions). In practice this results in a higher level of control over design evolution and simpler implementation of system improvements. Many system elements have gained in conceptual simplicity and consequent ease of management thanks to loose typing and the adoption of a unified approach to their online manipulation: activities/scripts and their methods; member types and instances; properties and primitives; items and collections; and outcome schemas and views. In forcing these into the data collection through activity execution mechanism, at different levels of abstraction, CRISTAL forces its meta model on all parts of the system, and with it comes a natural flexibility and evolvability. We hope to develop the modelling objects in CRISTAL in the future, instantiating concepts such as the use-case as Items and developing them into specification and annotation objects for future CRISTAL modules.

Research into the further extension and uses of CRISTAL continues. There are plans to enrich its kernel (the data model) to model not only data and processes (products and activities as items) but also to model agents and users of the system (whether human or computational). In the EC FP7 IAPP project 'CRISTAL-ISE', we are working with Agilium and an MES company, Alpha3i, to investigate how the semantics of CRISTAL items and agents could be captured in terms of ontologies and thus mapped onto or merged with existing ontologies for the benefit of new domain models. This project is also looking at improving the relationships between different CRISTAL nodes, to better handle pushing of new description versions from development to production, and propagating customer implemented changes back to the developers.

Following on from the work done in the neuroscience analysis domain, we see a potential to catalog and index any large archive of datafiles using any metadata available. CRISTAL can track all changes and accesses on those files, and their lifecycles can include verification and maintenance operations. XML is a very discoverable format, especially when accompanied by XML Schema, which serves as both validation and documentation of the data. We see potential applications of this in long-term data preservation, as we can guarantee that their data structures will be readable over long timescales as we can store the definitions of those structures alongside the data itself. We can also open up and add value to datasets for outreach, as the tools and understanding of manipulation of XML-based data is common amongst citizen scientists, while the data owners can measure the popularity of their published data elements.

**Acknowledgements**

The authors wish to thank their respective institutions for the support given during the development of the CRISTAL software. The long-scale development, testing and refinement of the CRISTAL design philosophy and model structure would not have been possible without the continued support of those institutions. In particular they would like to acknowledge the invaluable help and advice given to them by the following : Jean-Pierre Vialle, Thierry Le Flour, Sophie Lieunard, Alain Bazan and Steve Murray then of LAPP/CNRS, Annecy (France), Etiennette Auffray, Paul Lecoq and Guy Chevenier from CERN (Switzerland) and Zsolt Kovacs, Tony Solomonides, Florida Estrella and Norbert Toth from the University of the West of England (UK). They would also like to acknowledge the positive contributions from a number of external organisations that have allowed the exploitation of CRISTAL in practical commercial and research environments.